# A Systematic Literature Map on Big Data

Rogério Rossi, Kechi Hirama, Eduardo Ferreira Franco
Department of Computer and Digital System Engineering, Escola Politécnica of University of São Paulo, São Paulo, São Paulo, Brazil

***Abstract***— *The paradigm of Big Data has been established as a solid field of studies in many areas such as healthcare, science, transport, education, government services, among others. Despite widely discussed, there is no agreed definition about the paradigm although there are many concepts proposed by the academy and industry. This work aims to provide an analytical view of the studies conducted and published regarding the Big Data paradigm. The approach used is the systematic map of the literature, combining bibliometric analysis and content analysis to depict the panorama of research works, identifying patterns, trends, and gaps. The results indicate that there is still a long way to go, both in research and in concepts, such as building and defining adequate infrastructures and standards, to meet future challenges and for the paradigm to become effective and bring the expected benefits.*

## I. INTRODUCTION

Since, the mid-1980s the term Big Data has been used as a paradigm of high relevance for the academy, industry, executives from different business areas, and society in general who are involved wi.th modern components and digital information systems. As mentioned by [1], the Big Data era has reached several sectors, from government and e-commerce to healthcare organizations.

Respected and renowned research institutions, such as McKinsey & Company [2], The Data Warehouse Institute [3], IBM Research [4], among others, presented technical reports demonstrating the high relevance of studies related to Big Data. As there are divergences among studies and technical reports on the subject, it is not possible to verify a standard definition for Big Data allowing verifying different concepts for the term.

Besides its concepts, the properties associated with the term can be verified by many authors [4–8]. However, when its properties are considered, it is possible to identify a consensus associated with the Big Data 3V's Properties that corresponds to Volume, Variety, and Velocity.

Big Data is now presenting some relevant challenges in the field of technology infrastructure, the people involved, and the processes and business models associated with this digital information paradigm.

As a paradigm of large amplitude, it reaches organizations devoted to financial, commercial, and industrial business; social areas; governments, scientific associations and other environments that benefit from the information in the digital era. It also reaches universities to provide answers regarding to Big Data questions as also to efficiently and quickly prepare new qualified human resources.

Consequently, some bibliometric indicators presented results in technical and scientific publications related to Big Data [1,6]. In addition to the technical and scientific reports, consistent results are presented by the information industry [3], representing a continuous movement to meet the needs introduced by this new paradigm of computing.

There are many researches works in different scientific and social fields regarding to Big Data. The use and application of Big Data have also increased significantly according to the reports presented by IDC [9] and Gartner [10], which suggest more formal and deeper exploitation and considerations on this paradigm. Considering this scenario, the goal of this work is to provide an analytical view of the studies published on the subject in a highly relevant scientific database. To achieve this goal, this work conducted a systematic literature map, merging bibliometric and content analysis.

To achieve its goal, the paper is organized into five sections. Section two presents the research method adopted to collect and to analyze the sample of selected articles. Section three presents the results from the bibliometric and content analysis, followed by the discussion of the results in section four. Finally, section five presents the conclusions and limitations of this study and suggestions for future work.

## II. RESERCH DESIGN

The main objective of this work is to outline the scenario of the scientific literature on the "Big Data" topic using the bibliometric and content analysis of the selected sample articles, describing the trends, the key themes covered, identify the main journals that represent the discussion forum of the theme; who are the major authors; if there has been an increase in the number of publications over the period analyzed; which work most influenced the research community; and the main topics studied.

Aligned with the research objective the selected methodological approach was a systematic literature map, mixing bibliometric analysis [14] and content analysis [15], which are complementary [16].

The methodological approach selected for this study was the bibliometric analysis, which is a technique that allows assessing the existence of patterns in the literature on a particular topic, identifying the main journals dedicated to promoting and discussing a subject, the evolution of these publications over time, and what issues most relate to the searched subject. Furthermore, the approach enables to perform the analysis of the citations, which identify the works that most impacted the area by building the paper-paper citation network, and between those selected works and the most cited references [11,12].

Bibliometric analysis enables to perform a quantitative analysis of the citations between articles. This analysis can be made by counting the number of the individual citations of each paper, as well as the analysis of the references used by the most cited articles, allowing the researcher to identify bibliometric clustering phenomena and relationships between two articles based on the number of common references [13].

The scientific database "Web of Science" was used to collect the sample by the combination of the terms "("bigdata" OR "big data")" to search the titles of the publications, resulting in a set of 2,718 articles.

This database was selected because it provides an interface to simultaneously search across different sources using a common set of search fields for obtaining comprehensive results. It includes studies from 1985 to the current date, covering the Science Citation Index Expanded, Social Sciences Citation Index, Arts & Humanities Citation Index, and Emerging Sources Citation Index, which comprehends studies from ACM, EBSCOhost, Elsevier, Emerald, IEEE, INFORMS, ProQuest, SAGE, Springer, Taylor & Francis, Wiley, among many other publishers. This database is also the source for computing the "Journal Citation Report" index, which is one of the most used mechanisms for evaluating journals based on citation data.

From this initial set, only the works in the "Article" category were selected. This filtering was performed because, according to [17], these works go through a peer review process before being published and they also contain the information necessary for performing the bibliometric analysis.

Additionally, some articles were excluded from the sample. The criteria for selection were: the articles had to be published in "English" and classified in the categories related to "Computer Science", "Business Economics", "Engineering", "Telecommunication", "Governmental Law", "Information Science Library Science", "Science Technology Other Topics" "Health Care Services Sciences", "Social Sciences Other Topics", "Communication", "Mathematics", "Operation Research Management Science", "Medical Informatics" and "Public Administration". After this screening, a 391-article subset was obtained from the initial sample.

Next, the articles titles were read and classified separately; in addition, the individual evaluation results were consolidated. The items that showed divergence in classification, as to maintaining them or not for further analysis, were discussed until a consensus emerged. From this analysis, 103 items were excluded for not being associated with the theme, reducing the sample to 288 items.

Finally, the papers were analyzed in two stages: 1) bibliometric and, 2) content analysis and codification. Fig 1. shows the workflow carried out for analyzing the selected sample, adapted [17].

The Sci2Tool version 1.1 software [18] was used for the bibliometric analysis and for constructing the citation of the articles and the keyword networks. For analyzing the articles that most impacted the area, the articles to references network was elaborated, because it provides not only an overview of the most cited articles, but also of its most cited references, allowing identifying those which contributed to articulating the theoretical foundations of the area.

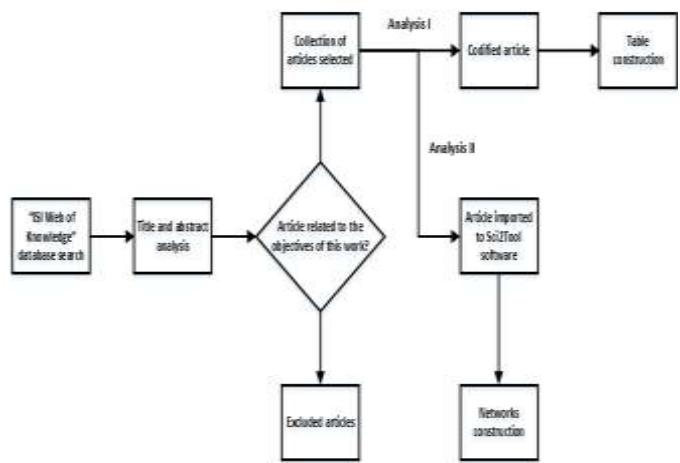

Fig. 1. Workflow conducted for analyzing the sample selected

The content analysis allows analytical flexibility for defining the codification scheme, which is then used in the code occurrence statistics and their relationships, as well as in the qualitative interpretation analysis [15]. Content analysis was used to identify the key issues and gaps in the literature, in addition to the type of research, the nature of the data and the affiliations of the authors.

For content analysis and codification, 11 articles were selected from the original sample of 288 items. The criteria for the selection was to extract the articles that received two or more citations from other articles in the sample. These articles were read and classified according to the coding scheme presented in Table I.

First, the articles were divided into two groups, according to the type of study: conceptual (theoretical/conceptual, modeling, literature review, and simulation) or empirical (survey, case study, experimental, action research, and technology). Then they were classified according to the affiliation of the authors; finally, they were evaluated as to the nature of the variables used in the research (quantitative and qualitative).

### III. RESULTS

#### A. Bibliometric Analysis

TABLE I. Code scheme used to classify selected works (adapted from [19]).

| C1 – Kind of study | C2 – Authors affiliation |
|---|---|
| TE1 – Modeling | A1 – University |
| TE2 – Theoretical / Conceptual | A2 – Research institution |
| TE3 – Literature review | A3 – Company |
| TE4 – Simulation | C3 – Approach |
| TE5 – Survey | ND1 – Quantitative |
| TE6 – Case study | ND2 – Qualitative |
| TE7 – Research action | |
| TE8 – Experimental | |
| TE9 – Technology | |

The 288 articles selected were published in 152 journals. Table II shows the number of publications per year, from 2009 and 2015, of the journals with at least four publications.

The first publications appeared in 2009, which shows that it corresponds to a recent research topic. Eight journals published approximately 19.5% of the sample: Computer, IBM Journal of Research and Development, Accounting Horizons, International Journal of Production and Economic, Software Practice & Experience, IEEE Network, International Journal of Communication and Communications of the ACM.

However, the vast majority of the selected articles, corresponding to approximately 67%, occurred in 143 different journals with three or fewer publications.

Among the journals shown in Table II, there are three of them that focus on areas beyond technology, which are: Accounting Horizons (targeting academic and professional audiences besides addressing issues related to accounting); International Journal of Production and Economic (interdisciplinary, focusing on the interface between engineering and management; also covering issues related to manufacturing, industrial processes and production); and MIT Sloan Management Review (discourses about advances in management practices that are transforming how people lead and innovate).

From the analysis of Table II, it is also possible to identify a growing trend in the number of publications over the years, ranging from 1 publication in 2009 to 116 in 2015.

TABLE II. Journal and period publication distribution

| JOURNAL | JCR | 2009 | 2011 | 2012 | 2013 | 2014 | 2015 | TOTAL |
|---|---|---|---|---|---|---|---|---|
| Computer | 1.443 | | | | 5 | 3 | 3 | 11 |
| IBM Journal of Research and Development | 0.688 | | | | 6 | 4 | 1 | 11 |
| Accounting Horizons | 0.881 | | | | | | 8 | 8 |
| International Journal of Production Economics | 2.752 | | | | | 1 | 5 | 6 |
| Software Practice & Experience | 0.897 | | | | 1 | | 4 | 5 |
| IEEE Network | 2.540 | | | | | 4 | 1 | 5 |
| International Journal of Communication | 0.618 | | | | | 1 | 4 | 5 |
| Communications of the ACM | 3.621 | 1 | | | 1 | 2 | 1 | 5 |
| Information Sciences | 4.038 | | | | | 3 | 1 | 4 |
| International Journal of Distributed Sensor Networks | 0.665 | | | | | 2 | 2 | 4 |
| Cluster Computing – The Journal of Networks Software Tools and Applications | 1.510 | | | | | | 4 | 4 |
| Future Generation Computer-Systems – The International Journal of Grid Computing and eScience | 2.786 | | | | | 3 | 1 | 4 |
| MIT Sloan Management Review | 1.529 | | 1 | 1 | | 1 | 1 | 4 |
| IT Professional | 0.819 | | | | 3 | | 1 | 4 |
| Journal of Internet Technology | 0.438 | | | | | 3 | 1 | 4 |
| Journal of Parallel and Distributed Computing | 1.179 | | | | | 1 | 3 | 4 |
| Journal of Systems and Software | 1.352 | | | | | | 4 | 4 |
| Knowledge-based Systems | 2.947 | | | | | 1 | 3 | 4 |
| Others | | 0 | 2 | 6 | 41 | 71 | 67 | 192 |
| TOTAL | | 1 | 3 | 7 | 58 | 103 | 116 | 288 |

A citation network was built to identify the articles of greater influence, in which each reference between two articles was represented as a link. After elaborating the paper citation network, the articles presenting at least two citations were selected and it is shown in Table III.

*B. Content Analysis and Coding*

The results shown in Table IV were obtained from the coding scheme (Table I) used to classify the selected works (Table III). The cells of Table IV indicate the presence of terms associated with the respective code in each of the 11 selected articles.

The following sub-sections discusses the relevant points of each of the 11 works presented, based on the coding scheme.

*University/Research Institution/Enterprise (A1-A2-A3)*

The articles selected showed the presence of a representative set of technical works related to Big Data applications. Most articles are from universities and research institutions (A1-3), with little association with enterprise works, which shows the Big Data paradigm is still new.

TABLE III. Most cited articles from the sample

| # | Title | Authors | # Citations |
|---|---|---|---|
| 1 | Business Intelligence and Analytics - From Big Data to Big Impact | [1] | 17 |
| 2 | Critical Questions for Big Data | [20] | 13 |
| 3 | The pathologies of big data | [21] | 10 |
| 4 | Data Science, Predictive Analytics, and Big Data - A Revolution That Will Transform Supply Chain Design and Management | [22] | 5 |
| 5 | Data-intensive applications, challenges, techniques and technologies - A survey on Big Data | [23] | 5 |
| 6 | Data mining with big data | [24] | 4 |
| 7 | The meaningful use of big data - four perspectives four challenges | [25] | 3 |
| 8 | Data quality for data science, predictive analytics, and big data in supply chain management - An introduction to the problem and suggestions for research and applications | [26] | 2 |
| 9 | A cubic framework for the chief data officer - Succeeding in a world of big data | [27] | 2 |
| 10 | Hazy - making it easier to build and maintain big-data analytics | [28] | 2 |
| 11 | Data quality management, data usage experience and acquisition intention of big data analytics | [29] | 2 |

TABLE IV. Code schema associated with selected works

| # | Article/Author | TE1 | TE2 | TE3 | TE4 | TE5 | TE6 | TE7 | TE8 | A1 | A2 | A3 | ND1 | ND2 |
|---|---|---|---|---|---|---|---|---|---|---|---|---|---|---|
| 1 | [1] | | | | | | | | | | | | | |
| 2 | [20] | | | | | | | | | | | | | |
| 3 | [21] | | | | | | | | | | | | | |
| 4 | [22] | | | | | | | | | | | | | |
| 5 | [23] | | | | | | | | | | | | | |
| 6 | [24] | | | | | | | | | | | | | |
| 7 | [25] | | | | | | | | | | | | | |
| 8 | [26] | | | | | | | | | | | | | |
| 9 | [27] | | | | | | | | | | | | | |
| 10 | [28] | | | | | | | | | | | | | |
| 11 | [29] | | | | | | | | | | | | | |

*Quantitative (ND1)*

It is also possible to notice that Analytics has been used since 2000 and recently has evolved into Big Data Analytics (ND1).

For example: [1] presented a trend of Business Intelligence and Analytics (BI&A) works between 2000-2011 with 3,602 works classified in Business Intelligence (3,146), Business Analytics (213), and Big Data (243). Business Analytics and Big Data started to emerge in 2007; 2011 was the most expressive year, with publications related to Business Intelligence (338), Business Analytics (126), and Big Data (95).

*Qualitative (ND2)*

As qualitative approach (ND2), [1] presented an overview of the impact that Big Data had in some Business Intelligence applications. For example, in "E-Commerce and Market Intelligence", the user data logs, records, and contents generated by customers can affect their faithfulness and satisfaction.

Reference [20] presented six issues with the emergence of Big Data that refer to new knowledge interpretation, subjective data analysis, the importance of systematic and rigorous approach to data collecting and analysis, Big Data contextualization, data accessibility versus ethics, public and private access.

[21] presented the pathologies of Big Data facing the appearance of good performance devices due to the increasing quantity of data.

Reference [22] presented the causes that affect Big Data volume, velocity and variety attributes. Among data types, for example, the sales department needs more details about its operations, including prices, dates, customers (volume) data, the monthly, weekly, daily or hourly (velocity) frequency, straight, distributors, internet and competitors (variety) sales.

Reference [23] discussed opportunities and challenges to US government sectors that believe in Big Data usefulness to decision making based on intensive data. The challenges regard data analysis of Big Data, taking into account inconsistencies, non-completeness, and scalability based on time and data security.

[24] treated Big Data as enormous data volume, heterogeneous, autonomous sources with a centralized and distributed control, which explores complex and evolved relationships. The challenges of Data Mining with Big Data include three layers: access and data computation; privacy and domain knowledge; and the mining algorithm of Big Data.

Reference [26] presented the data quality problem in the Supply Chain Management (SCM) context and proposed methods of monitoring and controlling data quality based on Statistical Process Control (SPC).

*Modeling (TE1)*

Among conducted studies, only one of the eleven selected articles presented a framework of Big Data processing [24] (TE1). The framework consists of three layers: 1) Big Data Mining Platform, which accesses and computes low-level data; 2) Information Sharing, Data Privacy, Big Data Applications and Knowledge, which deals with user privacy issues, high-level semantics and knowledge of application domain; and 3) Big Data Mining Algorithms, which processes the data mining.

*Theory (TE2)*

The selected works presented technologies and correlated research areas, issues that still require more definitions and research on using Big Data, a framework for processing Big Data, and methods of monitoring and controlling data quality (TE2).

[1] presented the foundational technologies, which are mature and well evolved, and some emerging researches being developed in Analytics. The Analytic types highlighted are (Big) Data Analytics, Text Analytics, Web Analytics, Network Analytics, and Mobile Analytics.

Reference [20] presented critical issues about Big Data that require more research and definitions for expanding its usage, such as Big Data changes the knowledge definition that requires new functionalities for searching and archiving; data interpretation that requires objectivity and accuracy; the rigorous and systematic way for collecting and analyzing data, independently of their size.

[24] presented the HACE theorem charactering Big Data as Heterogeneous, Autonomous, Complex and Evolved features, proposing a framework for Big Data processing from a Data Mining perspective in three layers (data accessing and computing, data privacy and domain knowledge, Big Data mining algorithms).

*Literature Review (TE3)*

Articles presenting a literature review (TE3) showed the general view of concepts, but according to [1], due to recent interests in Big Data, there is still no significant amount of work discussing the many tendencies. In this sense, the work of [1] aimed to analyze the impact that Business Intelligence & Analytics (BI&A) had on the currently growing importance of data in several critical areas including commerce, government, science and technology, health, public security and safety.

*Survey (TE5)*

Some authors presented research of predictions, challenges and opportunities of Big Data (TE5). For example, [1] presented some predictions about the future impacts of Big Data on some BI&A applications. For E-Commerce and Market Intelligence, the authors highlighted that long-tail marketing, personalized and directed recommendation, sales increase, and customer satisfaction are important issues.

Reference [23] presented the opportunities and challenges of Big Data. The USA government sectors believe in the usefulness of Big Data to decision making based on intensive data.

[29] presented a research based on the theory perspective of data management among 306 selected industries, with experience in using internal and external data, and which the effort in data quality revealed positive effect to Big Data Analytics adoption.

*Case Study (TE6)*

Reference [21] presented a case study (TE6) about some pathologies originated due to the large amount of data, which brings difficulties in using it. Other difficulties involve media storage to deal with huge datasets, besides the manipulated data size considering the analytical applications as the last stage of the data analysis. For dealing with these difficulties, the authors proposed the Distributed Computing as a strategy for Big Data.

*Experimental (TE8)*

We can note discussions about applications presented by some authors. For example, [1] discussed some applications of BI&A. In E-Commerce and Market Intelligence, examples are Recommender Systems, Social Media Monitoring and Analysis, Social and Virtual Games. In E-Government and Politics 2.0, Ubiquitous Government Services, Citizen Engagement and Participation and Political Campaign and e-Pooling are the most important.

Reference [22] presented some potential applications of Big Data in logistics and SCM. For example, in manufactures, fast response to the positive and negative customer sentiments, stocks managed by retailers, efficient response to customers, improvement of delivery tracking system regarding time and availability, and more effective monitoring of productivity. In Retail, mobile devices and customer data sentiments in stores, improvement of accuracy of stocks, linking of local traffic and climate, work reduction due to stock error reductions.

[28] treated the GeoDeepDive, a demonstration project that illustrates Hazy approach to the trained systems development. The GeoDeepDive project had the support of geologists to do the linguistics and statistics analysis on 10-ton of journal articles in geology.

IV. DISCUSSION

As part of the results, this section addresses some concepts based on the most cited keywords presented in the eleven most cited articles.

From the extensive set of keywords identified, some of them, which have their frequency index greater than three, were selected for the analysis as shown in Table V. It is also presented its relationship with the eleven most cited articles presented in Table III.

The first keyword, i.e., the most mentioned by all articles corresponds to *Big Data*. The considerations related to the other keywords are presented in an analytical and conceptual view according to the most cited word, i.e., *Big Data*.

As can be seen in Table V, the keyword *Big Data Analytics* was mentioned three times by considering the set of eleven articles, representing the most mentioned keyword. All the others were cited only once.

With the set of keywords identified, it is possible to verify that there is a close relationship between some of them, promoting an integrated approach to specific concepts.

*Big Data Analytics*, *Analytics*, and *Predictive Analytics* are the keywords frequently cited offering prominence to define a term that has been associated with Big Data, which refers to Analytics or also to Big Data Analytics. The terms Big Data and Analytics have different definitions, sometimes interchangeable and sometimes generating conceptual conflict, as can be seen in [30], describing that some executives are questioning whether Big Data is not just another way of saying Analytics.

TABLE V. Keywords with frequency index greater than 3 identified in the eleven most cited articles

| Keywords | Frequency index | Selected articles | | | | | | | | | | |
|---|---|---|---|---|---|---|---|---|---|---|---|---|
| | | 1 | 2 | 3 | 4 | 5 | 6 | 7 | 8 | 9 | 10 | 11 |
| *Big Data* | 115 | | | | | | | | | | | |
| *Cloud Computing* | 17 | | | | | ▓ | | | | | | |
| *MapReduce* | 12 | | | | | | | | | | | |
| *Data Mining* | 9 | | | | | | ▓ | | | | | |
| *Hadoop* | 9 | | | | | | | | | | | |
| *Big Data Analytics* | 8 | ▓ | | | | | | | | | ▓ | ▓ |
| *Analytics* | 7 | | ▓ | | | | | | | | | |
| *Social Media* | 6 | | ▓ | | | | | | | | | |
| *Data Analytics* | 5 | | | | | | | | | | | |
| *Predictive Analytics* | 5 | | | | | ▓ | | | | | | |
| *Machine Learning* | 5 | | | | | | | | | | ▓ | |
| *Text Mining* | 5 | | | | | | | | | | | |
| *Internet of Things* | 4 | | | | | | | | | | | |
| *Business Intelligence* | 4 | ▓ | | | | | | | | | | |
| *Computational Social Science* | 4 | | | | | | | | | | | |

The notion of Analytics is quite variable and it is possible to verify some definitions for this term generally referring to the extraction of knowledge from information. To [1] Analytics associated with Big Data is a term that has been used to describe the data sets and techniques in large and complex applications demanding advanced storage, management, analysis, and visualization technologies.

For more specific cases it is possible to identify the term Analytics associated with Big Data, creating the term Big Data Analytics. It is also possible to identify some definitions for this term, which is emphasized by [29] as the technologies and techniques that a company can employ to analyze large-scale, complex data, for various applications intended to augment firm performance in various dimensions.

To [22] Predictive Analytics is a subset of data science having a number of disciplines related to this, such as Statistics, Forecasting, Optimization, Applied Probability, Data Mining, and Analytical Mathematical Modeling. Predictive Analytics comprise a variety of techniques that predict future outcomes based on the historical and current data.

It is relevant to the current scenario involving analytics and Big Data to present the classification proposed by [1] for emerging research on analytics associated with Big Data. The authors proposed five technical categories for analytics and Big Data: 1) (Big) Data Analytics, 2) Text Analytics, 3) Web Analytics, 4) Network Analytics, and 5) Mobile Analytics. It is also possible to verify in [5] other categories that were highlighted for analytics related to Big Data: 1) Structured

Analytics, 2) Text Analytics, 3) Web Analytics, 4) Multimedia Analytics, 5) Network Analytics, and 6) Mobile Analytics.

The concept related to the analytics has been presenting significant challenges when it comes to Big Data. Perhaps it is one of the most important challenges for this paradigm, as it refers to the semantic capacity of Big Data.

Although different challenges are checked for the different organizations that need integrated Analytics to Big Data, [31] commented that the main barriers faced by organizations that adopt analytics with Big Data are the management and cultural, besides, they are not necessarily linked to data and technologies. For [32] the value of any analytics is largely related to the capacity that it offers to decision makers.

From the integrated presentation of the keywords highlighted above related to the analytics concept, the other three keywords that have high relevance and are also highlighted by researchers and the industry involved in Big Data are: *Machine Learning*, *Data Mining*, and *Business Intelligence*.

It should be noted that these three keywords mentioned in several articles are words that refer to broad concepts, so this article provides the relationship of these keywords by addressing them conceptually and by analyzing the integration that can be verified from these concepts.

The analysis from these other three keywords has the objective of presenting their main features, concepts, and especially the relationship of them with Big Data.

The principle of *Data Mining* is to extract knowledge from information using algorithms, models, platforms, and specific technologies to meet this objective related to data management.

According to [1] and [33], some of the most influential Data Mining algorithms refer to C4.5, k-means, SVM, Apriori, EM (Expectation Maximization), PageRank, AdaBoost, kNN (k-Nearest neighbors), Naive Bayes, and CART covering problems of classification, clustering, regression, association analysis, and network analysis. However, these do not meet all the needs of Big Data, representing a clear technical barrier for this paradigm.

Mining operation in Big Data, as cited by [1], culminates in studies that turn to analytics to meet the shares of classification, prediction, regression, association rule, and clustering analysis where machine learning and genetic algorithms have been contributing to the success of different data mining applications.

To [23] there is an integration of data analysis techniques involving Data Mining, Neural Networks, Machine Learning, Signal Processing, and Visualization Methods. It is possible to note that Machine Learning practices, Statistical Theories and Models, and Multivariate Analysis represent a relevant data analysis technique that culminates in Data Mining. Therefore, it can be concluded that the Statistical Models collaborate to the application of Machine Learning Methods in Data Mining tasks favoring the activities of knowledge discovery considering Big Data.

As examples, there are some technologies that address the integration of practices and methods of Machine Learning and Data Mining applied to Big Data, such as: Hazy Project [28]; MLBase [34]; Weka [35]; and Stratrosphere [36].

As mentioned by Chen et al. [1], Business Intelligence with Analytics corresponds to Big Data as an important area, either in the academic or corporate environments.

Reference [1] propose the unification of Business Intelligence and Analytics addressing questions about Big Data Analytics as a way to offer new directions to Business Intelligence and Analytics. They consider the evolution of Business Intelligence and Analytics standing as follows: BI&A 1.0 with focus to structured data (DBMS-based, content structured); BI&A 2.0 considering the Internet as a new possibility for data collection and the advent of web analytics tools (Google Analytics) (Web-based, unstructured content); and BI&A 3.0 which covers the era of large-scale use of mobile phones and the beginning of the actions related to the new paradigm that corresponds to the Internet of Things (Mobile and Sensor-based content).

Considering another keyword highlighted in Table V, there is the concept of virtualization technologies culminating in one of the most robust technologies for Big Data corresponding to *Cloud Computing*, which together with other technologies has become a mechanism capable of offering effective response to the needs presented in a petabyte-scale data as [23]. To [37] Cloud Computing is closely linked to Big Data and should be used for the management of huge computing and storage resources providing computing capacity for Big Data Applications. As complemented by [38], cloud maintain more than the hardware, it gives to customers a set of virtual machines for Big Data Management.

As another form of high-volume data generation, *Social Media* is linked to Big Data becoming an important instrument for decision-making in many sectors. Considering text messages exchanged by people, e-mail messages, and mechanisms for providing images, Social Media has reached more and more users. As [20], Social Media interaction is part of behavioral networks.

Another relevant point related to Social Media is presented by [1] referred to the creation of many techniques for analysis of user opinion, text, and sentiment analysis. These types of analysis of online community behavior and social network culminate in what the authors call Network Analytics including new techniques and computational models.

Two relevants keywords related to Big Data are: *Hadoop* and *MapReduce*. Both have its frequency index greater than three, but they were not mentioned by the authors of the eleven most cited articles, so, as they have high impact and relevance for Big Data paradigm, they will be treated as follow.

*Hadoop* supports massive data storage and has high scalability and processing power distributed to deal with a huge amount of data and corresponds to an open-source framework. Being used with NoSQL databases it can provide flexibility for Big Data Applications [39]. Hadoop is also regarded as the best software platform established to support high data volume implementing a distributed computing paradigm called *MapReduce* as [23].

Based on the divide and conquer method, the computational model *MapReduce*, developed by Yahoo! and other web companies, treat a complex problem through several sub-problems using the *map* step and *reduce* step. The solution is ensured by the combination of the sub-problems solutions, which has led the Hadoop/MapReduce model to a wide use for Big Data.

MapReduce as a programming model operates with two basic functions that refer to map and reduce. Moreover, as part of the Apache Hadoop, it has emerged as a very efficient model for Big Data processing as [40].

Thus, the concepts associated with the set of keywords that are involved with Big Data were covered up, allowing a better understanding of this paradigm. Similarly, this approach can foster future research able to define and structure the concepts and applications related to Big Data to offer better results.

## V. CONCLUSION

This work aimed to depict the panorama of published studies regarding to Big Data evaluating its maturity by following a consistent method of scientific research. For analyzing the published scientific articles, the "Web of Science" (WoS) database was used along with the bibliometric analysis following a specific workflow to favor the observation of the production and expansion of scientific articles considering Big Data.

The bibliometric analysis allowed some detailed responses presented in two formats: the first one from the perspective of some categories presented in Table IV and the second one based on the concepts observed from the keywords with higher frequency index and presented in the eleven most cited articles as Table V.

A detailed review and analysis of these studies allowed observing that this is a term involved in many research areas and presented in a variety of complexity levels.

Big Data scientific studies concern areas such as economics, business, healthcare, information systems, among others. Some research works are related to new technologies, processes and models addressing the Big Data Architecture and Infrastructure as verified in [28] and [36]; as also some are linked to expectations considering Big Data responses.

Thus, since the works have varying levels of complexity, it was possible to find articles with high scientific formalism, besides works that present certain superficiality regarding results. Basic concepts observable in most studies are greatly repeated, which leads to a sense of solidification of this initial stage of research.

There are some surveys on Big Data, many specific works related to a particular subject or area, and others that involve technologies to support Big Data. However, a slow convergence of these studies or surveys was observed; again, it is possible to note that although linearly the research works present some advance, in its integrative way, they do not converge.

Embracing challenges and powerful results, the Big Data paradigm has been established in many areas, despite not presenting concrete and replicable results yet, but as a solid element of studies and research that may provide significant improvements in sectors such as healthcare, science, transport, education, government services, among others.

ACKNOWLEDGMENT

We thank Programa de Educação Continuada em Engenharia (PECE) of Polytechnic School of University of São Paulo for supporting this work.

REFERENCES

[1] H. Chen, R. H. L. Chiang, V. C. Storey, "Business intelligence and analytics: from big data to big impact," *MIS Quarterly*, vol. 36. pp.1165–1188, 2012.
[2] J. Manyika, M. Chui, B. Brown, J. Bughin, R. Dobbs, C. Roxburgh, A.H. Byers, "Big data: The next frontier for innovation, competition, and productivity," McKinsey Global Institute, 2011. http://www.mckinsey.com/business-functions/business-technology/our-insights/big-data-the-next-frontier-for-innovation.
[3] P. Russom, "Managing Big Data," 2013. http://tdwi.org/webcasts/2013/10/managing-big-data.aspx.
[4] M. Schroeck, R. Shockley, J. Smart, D. Romero-Morales, P. Tufano, "Analytics: The real-world use of big data," *IBM Global Business Services*, vol. 12, pp. 1-20, 2012.
[5] H. Hu, Y. Wen, T. Chua, X. Li, "Toward Scalable Systems for Big Data Analytics: A Technology Tutorial," *IEEE Access*, vol. 2, pp.652–687, 2014. doi:10.1109/ACCESS.2014.2332453.
[6] R. T. Bedeley, L. S. Iyer, "Big Data Opportunities and Challenges: The Case of Banking Industry", in *Proceedings of the Southern Association for Information Systems Conference*, Macon, GA, USA, pp. 1–6, 2014.
[7] NIST, Draft NIST Big Data Interoperability Framework: Volume 1, 2015.
[8] Y. Demchenko, C. de Laat, P. Membrey, "Defining architecture components of the Big Data Ecosystem," in *2014 International Conference on Collaboration Technologies and Systems (CTS) IEEE*, 2014. doi:10.1109/CTS.2014.6867550.
[9] IDC, "Worldwide Big Data Technology and Services forecast 2016-2020," 2016. https://www.idc.com/getdoc.jsp?containerId=US40803116.
[10] Gartner, "Gartner Survey Shows More Than 75 Percent of Companies Are Investing or Planning to Invest in Big Data in the Next Two Years," 2015. http://www.gartner.com/newsroom/id/3130817.
[11] A. Neely, "The evolution of performance measurement research: Developments in the last decade and a research agenda for the next," *International Journal of Operations & Production Management*, vol. 25, no. 12, pp.1264–1277, 2005. doi:10.1108/01443570510633648.
[12] S. Prasad, J. Tata, "Publication patterns concerning the role of teams/groups in the information systems literature from 1990 to 1999," *Information & Management*, vol. 42, no. 8, pp.1137–1148, 2005. doi:10.1016/j.im.2005.01.003.
[13] M. M. Kessler, "Bibliographic coupling between scientific papers," *American Documentation*, vol. 14, no. 1, pp. 10- 25, 1963. doi:10.1002/asi.5090140103.
[14] L. Ikpaadhindi, "An overview of bibliometric: its measurements, laws and ther application," *Libri*, vol. 35, 1985.
[15] V. J. Duriau, R. K. Reger, M. D. Pfarrer, "A content analysis of the content analysis literature in organization studies: Research themes, data sources, and methodological refinements," *Organizational Research Methods*, vol. 10, no. 1, pp.5–34, 2007. doi:10.1177/1094428106289252.
[16] M. M. Carvalho, A. Fleury, A. P. Lopes, "An overview of the literature on technology roadmapping (TRM): Contributions and trends," *Technological Forecasting and Social Change*, vol. 80, no. 7, pp.1418–1437, 2013. doi:10.1016/j.techfore.2012.11.008.
[17] M. M. Carvalho, P. Lopes, D. Marzagão, "Gestão de portfólio de projetos: contribuições e tendências da literatura," *Gestão & Produção*, vol. 20, pp. 433–453, 2013.
[18] Sci2 Team, Science of Science (Sci2) Tool, 2009. https://sci2.cns.iu.edu
[19] J. A. Carnevalli, P. C. Miguel, "Review, analysis and classification of the literature on QFD-Types of research, difficulties and benefits," *International Journal of Production Economics*, vol. 114, no. 2, pp.737–754, 2008. doi:10.1016/j.ijpe.2008.03.006.

[20] D. Boyd, K. Crawford, "Critical questions for big data: Provocations for a cultural, technological, and scholarly phenomenon," *Information, Communication and Society*, vol. 15, no. 5, pp.662–679, 2012. doi:10.1080/1369118X.2012.678878.
[21] A. Jacobs, "The pathologies of big data," *Communications of the ACM*, vol. 52, no. 8, pp. 36-44, 2009. doi:10.1145/1536616.1536632.
[22] M. A. Waller, S. E. Fawcett, "Data science, predictive analytics, and big data: A revolution that will transform supply chain design and management," *Journal of Business Logistics*, vol. 34, pp.77–84, 2013. doi:10.1111/jbl.12010.
[23] C. P. Chen, C. Y. Zhang, "Data-intensive applications, challenges, techniques and technologies: A survey on Big Data," *Information Sciences*, vol. 275, pp. 314–347, 2014. doi:10.1016/j.ins.2014.01.015.
[24] X. Wu, X. Zhu, G. Q. Wu, W. Ding, "Data mining with big data," *IEEE Transactions Knowledge and Data Engineering*, vol. 26, no. 1, pp. 97–107, 2014. doi:10.1109/TKDE.2013.109.
[25] C. Bizer, P. A. Boncz, E. L. Brodie, O. Erling, "The meaningful use of big data," *ACM SIGMOD Record*, vol. 40, pp. 56-60, 2011. doi:10.1145/2094114.2094129.
[26] B. T. Hazen, C. A. Boone, J. D. Ezell, L. A. Jones-Farmer, "Data quality for data science, predictive analytics, and big data in supply chain management: An introduction to the problem and suggestions for research and applications," *International Journal of Production Economics*, vol. 154, pp.72–80, 2014. doi:10.1016/j.ijpe.2014.04.018.
[27] Y. Lee, S. Madnick, R. Wang, F. Wang, H. Zhang, "A cubic framework for the chief data officer: Succeeding in a world of big data," *MIS Quarterly Executive*, vol. 13, pp.1–13, 2014.
[28] A. Kumar, F. Niu, C. Ré, "Hazy: making it easier to build and maintain big-data analytics," *Communications of the ACM*, vol. 56, no. 3, pp.40-49, 2013. doi:10.1145/2428556.2428570.
[29] O. Kwon, N. Lee, B. Shin, "Data quality management, data usage experience and acquisition intention of big data analytics," International Journal of Information Management, vol. 34, no. 3, pp. 387–394, 2014. doi:10.1016/j.ijinfomgt.2014.02.002.
[30] A. McAfee, E. Brynjolfsson, "Big data: The management revolution," *Harvard Business Review*, vol. 90, no.10, pp. 60–68, 2012.
[31] S. LaValle, M. S. Hopkins, E. Lesser, R. Shockley, N. Kruschwitz, "Analytics: The new path to value," *MIT Sloan Management Review*, vol. 52, pp.1–25, 2014.
[32] D. Kiron, P. K. Prentice, R. B. Ferguson, "Raising the bar with Analytics," *MIT Sloan Management Review*, vol. 55, no. 2, 2014.
[33] X. Wu, V. Kumar, J. R. Quinlan, J. Ghosh, Q. Yang, H. Motoda, G.J. McLachlan, A. Ng, B. Liu, P. S. Yu, Z. H. Zhou, M. Steinbach, D. J. Hand, D. Steinberg, "Top 10 algorithms in data mining," *Knowledge and Information Systems*, vol. 14, no.1, pp.1–37, 2008. doi:10.1007/s10115-007-0114-2.
[34] T. Kraska, A. Talwalkar, J. Duchi, R. Griffith, M.J. Franklin, M. Jordan, "MLbase: A distributed machine-learning system," in *6th Biennial Conference on Innovative Data Systems Research (CIDR'13)*, Asilomar, California, USA, 2013.
[35] M. Hall, E. Frank, G. Holmes, B. Pfahringer, P. Reutemann, I. H. Witten, "The WEKA data mining software," *ACM SIGKDD Explorations Newsletter*, vol. 11, no. 1, pp. 10-18, 2009. doi:10.1145/1656274.1656278.
[36] A. Alexandrov, R. Bergmann, S. Ewen, J. C. Freytag, F. Hueske, A. Heise, O. Kao, M. Leich, U. Leser, V. Markl, F. Naumann, M. Peters, A. Rheinländer, M. J. Sax, S. Schelter, M. Höger, K. Tzoumas, D. Warneke, "The Stratosphere platform for big data analytics," *VLDB Journal*, vol. 23, pp.939–964, 2014. doi:10.1007/s00778-014-0357-y.
[37] M. Chen, S. Mao, Y. Liu, "Big Data: A Survey," *Mobile Networks and Applications*, vol. 19, no. 2, pp.171–209, 2014. doi:10.1007/s11036-013-0489-0.
[38] H. Demirkan, D. Delen, "Leveraging the capabilities of service-oriented decision support systems: Putting analytics and big data in cloud," *Decision Support Systems*, vol. 55, no. 1, pp. 412–421, 2013. doi:10.1016/j.dss.2012.05.048.
[39] J. O. Chan, "An architecture for big data analytics," *Communications of IIMA*, vol. 13, no. 2, 2013.
[40] S. Sharma, U. S. Tim, J. Wong, S. Gadia, S. Sharma, "A brief review on leading big data models," *Data Science Journal*, vol. 13, pp. 138–157, 2014. doi:10.2481/dsj.14-041.